\documentclass[english]{article}
\usepackage[T1]{fontenc}
\usepackage[latin9]{inputenc}
\usepackage{amssymb}
\usepackage{esint}

\makeatletter
\newcommand{\lyxaddress}[1]{
	\par {\raggedright #1
	\vspace{1.4em}
	\noindent\par}
}

\makeatother

\usepackage{babel}
\begin{document}
\title{\textbf{Universality of the thermodynamics of a quantum-mechanically
radiating black hole departing from thermality }}
\author{\textbf{$^{1,2,3}$Christian Corda and $^{4,1}$Carlo Cafaro}}
\maketitle

\lyxaddress{\textbf{$^{1}$SUNY Polytechnic Institute, 13502 Utica, New York,
USA }}

\lyxaddress{\textbf{$^{2}$Università eCampus, 22060 Novedrate (CO), Lombardy,
Italy }}

\lyxaddress{\textbf{$^{3}$International Institute for Applicable Mathematics
and Information Sciences, B. M. Birla Science Centre, Adarshnagar,
Hyderabad 500063, India }}

\lyxaddress{\textbf{$^{4}$University at Albany-SUNY, 12222 Albany, New York,
USA}}

\lyxaddress{\textbf{E-mails: }\textbf{\emph{cordac.galilei@gmail.com, ccafaro@albany.edu }}}
\begin{abstract}
Mathur and Mehta won the third prize in the 2023 Gravity Research
Foundation Essay Competition for proving the universality of black
hole (BH) thermodynamics. Specifically, they demonstrated that any
Extremely Compact Object (ECO) must have the same BH thermodynamic
properties regardless of whether or not the ECO possesses an event
horizon. The result is remarkable, but it was obtained under the approximation
according to which the BH emission spectrum has an exactly thermal
character. In fact, strong arguments based on energy conservation
and BH back reaction imply that the spectrum of the Hawking radiation
cannot be exactly thermal. In this work the result of Mathur and Mehta
will be extended to the case where the radiation spectrum is not exactly
thermal using the concept of \emph{BH dynamical state}.
\end{abstract}
In 1975 Hawking \cite{key-1} discovered that a Schwarzschild BH having
mass $M$ radiates at the Hawking temperature (hereafter Planck units
will be used, i.e. $G=c=k_{B}=\hbar=\frac{1}{4\pi\epsilon_{0}}=1$)
\begin{equation}
T_{H}=\frac{1}{8\pi M},\label{eq: Hawking temperature}
\end{equation}
with a corresponding probability of emission 
\begin{equation}
\Gamma\sim\exp(-\frac{\omega}{T_{H}}),\label{eq: hawking probability}
\end{equation}
where $\omega$ is the energy-frequency of the outgoing particle.
Hawking's approach was based on a previous work by Bekenstein \cite{key-2}
who associated the BH with an entropy given by 
\begin{equation}
S_{BH}=\frac{A}{4}=\pi r_{s}^{2},\label{eq: BH entropy}
\end{equation}
where $A$ is the horizon area and $r_{s}=2M$ is the Schwarzschild
radius. These pioneering results of Bekenstein and Hawking led to
the creation of the fascinating framework of black hole thermodynamics,
which is still the subject of many studies and investigations by theoretical
physicists from all over the world. In 1976, however, Hawking showed
that this approach led to a non-trivial problem \cite{key-3}, what
would later be known as the BH information paradox \cite{key-4,key-26}.
Hawking radiation is indeed created by the extreme force of gravity
present in the vacuum around the BH horizon. But this type of radiance
that is created by the quantum vacuum is not capable of carrying any
physical information with it. Thus, it would seem that even if one
solves the problem by showing that the BH actually has a different
physical structure, i.e. the structure of a normal body which emits
radiation from its surface like any other body, one would still be
left with the problem of explaining why this body should have the
BH thermodynamic properties one expects. This problem has been solved
in the remarkable paper of Mathur and Mehta \cite{key-4}. They showed
that any astrophysical ECO having a radius sufficiently close to the
Schwarzschild radius will have the same thermodynamic properties as
the semiclassical BH considered by Bekenstein and Hawking. On the
other hand, the analysis in \cite{key-4} starts with the assumption
that the Hawking radiation spectrum should be strictly thermal. However,
strong arguments based on energy conservation and BH back reaction
imply that the spectrum of BH radiance cannot be exactly thermal \cite{key-5}.
In fact, by considering the energy conservation, which means that
the BH contracts during particle emission, implying a varying BH geometry,
the important correction term \cite{key-5}
\begin{equation}
\Gamma\sim\exp[-\frac{\omega}{T_{H}}(1-\frac{\omega}{2M})]\label{eq: Parikh Correction}
\end{equation}
is obtained. In Eq. (\ref{eq: Parikh Correction}) the additional
term $\frac{\omega}{2M}\:$ is present. By introducing the \emph{effective
temperature }\cite{key-6}
\begin{equation}
T_{E}(\omega)\equiv\frac{2M}{2M-\omega}T_{H}=\frac{1}{4\pi(2M-\omega)},\label{eq: Corda Temperature}
\end{equation}
one can rewrite Eq. (\ref{eq: Parikh Correction}) in a Boltzmann-like
form similar to Eq. (\ref{eq: hawking probability}) 
\begin{equation}
\Gamma\sim\exp[-\beta_{E}(\omega)\omega]=\exp(-\frac{\omega}{T_{E}(\omega)}).\label{eq: Corda Probability}
\end{equation}
Here $\exp[-\beta_{E}(\omega)\omega]$ is the \emph{effective Boltzmann
factor,} where 
\begin{equation}
\beta_{E}(\omega)=\frac{1}{T_{E}(\omega)}=\beta_{H}\left(1-\frac{\omega}{2M}\right)\label{eq:  effective Boltzmann factor}
\end{equation}
with $\beta_{H}=\frac{1}{T_{H}}.$ Thus, in order to take into due
account the BH dynamical state given by the BH varying geometry during
particle emission, one must replace $T_{H}$ with $T_{E}(\omega)$
in the expression of the probability of emission. One must stress
that this happens also in several other fields of Science \cite{key-7}
where the deviation from the thermal spectrum of the emitting body
is usually taken into account via the introduction of an effective
temperature which represents the temperature of a black body emitting
exactly the same amount of radiation\emph{ }as the non-thermal source.
$T_{E}(\omega)$ is a function of the energy-frequency of the emitted
radiance and one can quantify the deviation of the radiation spectrum
from the strict thermality via the ratio $\frac{T_{E}(\omega)}{T_{H}}=\frac{2M}{2M-\omega}=1+\frac{\omega}{2M}+\mathcal{O}\left(\left|\frac{\omega}{2M}\right|^{2}\right)$.
One can also define other dynamical quantities like the BH\emph{ effective
mass and effective horizon radius }as \cite{key-6},
\begin{equation}
M_{E}\equiv M-\frac{\omega}{2}\quad and\quad\mbox{ }r_{E}\equiv2M_{E}.\label{eq: effective quantities emission}
\end{equation}
$M_{E}$ and $r_{E}$ are average quantities. Indeed, \emph{$r_{E}$
}is the average of the initial and final horizons, while \emph{$M_{E}$
}is the average of the initial and final masses \cite{key-6}. They
are the dynamical BH mass and horizon\emph{ during} the BH contraction,
i.e. \emph{during} the emission of the particle. This implies that
the analysis in \cite{key-4} can be improved with the replacement
$T_{H}$ $\rightarrow$$T_{E}(\omega)$ in the discussion. Of course,
despite the fact that this is very intuitive, it must be rigorously
justified. This can be done via the Gibbons-Hawking periodicity argument
\cite{key-8}. Following \cite{key-9}, the standard Schwartzschild
line-element is 
\begin{equation}
ds^{2}=-\left(1-\frac{2M}{r}\right)dt^{2}+r^{2}\left(\sin^{2}(\theta)d\varphi^{2}+d\theta^{2}\right)+\frac{dr^{2}}{1-\frac{2M}{r}}.\label{eq: Hilbert}
\end{equation}
Setting $\tau\equiv it$ and $x\equiv4M\sqrt{1-\frac{2M}{r}}$ \cite{key-10},
the Gibbons-Hawking argument {[}8--11{]} permits to write down the
Euclidean form of the metric in Eq. (\ref{eq: Hilbert}) as 
\begin{equation}
ds_{e}^{2}=x^{2}\left[\frac{d\tau}{4M\left(1-\frac{\omega}{2M}\right)}\right]^{2}+\left(\frac{r}{r_{E}}\right)^{2}dx^{2}+r^{2}\left(\sin^{2}(\theta)d\varphi^{2}+d\theta^{2}\right).\label{eq: Euclidean}
\end{equation}
In Refs. \cite{key-9,key-10}, the corrections to the standard semiclassical
Hawking temperature due to the quantum back reaction are estimated
by means of the Hamilton-Jacobi method. Specifically, the (complex)
semiclassical wave function of a massless particle in spacetime satisfying
the Klein-Gordon equation is expressed in terms of a (real) Hamilton-Jacobi
function (proportional to the phase of the complex wave function)
expanded in powers of $\hslash$. Then, terms from $O(\hslash)$ onwards
are treated as quantum corrections over the semiclassical value of
the Hamilton-Jacobi function. Within this theoretical scheme, the
corrected expression of the BH temperature is $T_{h}=T_{H}\left[1+\sum_{i}\beta_{i}\frac{\hbar^{i}}{M^{2i}}\right]^{-1},$
where $T_{H}$ is the standard semiclassical BH Hawking temperature
of Eq. (\ref{eq: Hawking temperature}). Note that $\beta_{i}$ are
dimensionless constant parameters and $\beta_{i}\frac{\hbar^{i}}{M^{2i}}$
is an adimensional quantity as well since the reduced Planck con$ $stant
has dimensions equal to the square of a mass. This is a consequence
of the fact that units $G=c=k_{B}=1$ are used in Refs. \cite{key-9,key-10}.
The Euclidean form of the metric of Eq. (\ref{eq: Euclidean}) is
regular at $x=0$ and $r=r_{E}$, while $\tau$ is considered as being
an angular variable with period $\beta_{E}(\omega)$ \cite{key-8,key-9}
. Now, if one replaces the quantity $\sum_{i}\beta_{i}\frac{\hbar^{i}}{M^{2i}}$
in \cite{key-9} with the quantity $-\frac{\omega}{2M}$as in Eq.
(\ref{eq: Parikh Correction}), by following step by step the analysis
in \cite{key-9} one gets the \emph{dynamical Schwartzschild metric}
as 
\begin{equation}
ds^{2}=-\left(1-\frac{2M_{E}}{r}\right)dt^{2}+r^{2}\left(\sin^{2}(\theta)d\varphi^{2}+d\theta^{2}\right)+\frac{dr^{2}}{1-\frac{2M_{E}}{r}},\label{eq: Hilbert dynamical}
\end{equation}
with the \emph{dynamical surface gravity }equal to $\frac{1}{4M_{E}}.$
Using Eq. (\ref{eq: effective quantities emission}) one can also
rewrite Eq. (\ref{eq: Hilbert dynamical}) as 
\begin{equation}
ds^{2}=-\left[1-\frac{2\left(M-\frac{\omega}{2}\right)}{r}\right]dt^{2}+r^{2}\left(\sin^{2}(\theta)d\varphi^{2}+d\theta^{2}\right)+\frac{dr^{2}}{1-\frac{2\left(M-\frac{\omega}{2}\right)}{r}}.\label{eq: Hilbert dynamical esplicita}
\end{equation}

This result implies that the analysis in \cite{key-4} can be improved
by using the dynamical quantities introduced here. We note that also
the entropy in Eq. (\ref{eq: BH entropy}) must be replaced with a
dynamical (effective) entropy 
\begin{equation}
\left(S_{BH}\right)_{E}\equiv\frac{A_{E}}{4}=\pi r_{E}^{2},\label{eq: dynamical BH entropy}
\end{equation}
where $A_{E}\equiv4\pi r_{E}^{2}$ is the \emph{BH effective horizon.
}An ECO was defined in \cite{key-4} via four properties which here
will be modified by taking into account the dynamical Schwarzschild
geometry which has been introduced above:
\begin{enumerate}
\item The ECO dynamical mass as seen from infinity is $M_{E},$ the Schwarzschild
BH dynamical mass as in Eq. (\ref{eq: effective quantities emission}). 
\item Denoting $R_{E(ECO)}$ the dynamical radius of the ECO, semiclassical
physics is a good approximation to the dynamics at $r\geq R_{E(ECO)}.$
\item The redshift in the semiclassical region reaches a value which is
the same as that of the redshift at distances of order of the Planck
length outside the dynamical horizon of the Schwarzschild BH. In that
case, the ECO is \emph{extremely compact}, in the sense that the dynamical
mass contained in the region $r\leq R_{E(ECO)}$ is approximately
$M_{E}-O(M_{E}).$
\item The theory governing the physical properties of the ECO does not violate
causality, so that, for larger values of $r\geq R_{E(ECO)}$, where
semiclassical physics holds, one has 
\begin{equation}
\frac{2M_{E}}{r}<1.\label{eq: Casualit=0000E0}
\end{equation}
\end{enumerate}
The analysis in \cite{key-4} is partially based on the previous result
\cite{key-12}. In that work the authors showed that if the ECO being
considered was a thin spherical shell supported by its own pressure,
and located just outside the horizon, then such a thin spherical shell
will be in thermodynamic equilibrium with the local Unruh radiation.
This will lead to the Bekenstein-Hawking entropy (\ref{eq: BH entropy})
for the shell. It is worth remembering that, historically, the first
to argue that the final state of gravitational collapse was a thin
and compact spherical shell of matter was Einstein in 1939 \cite{key-13}.
This is consistent with a notable result obtained by Vaz in 2014 \cite{key-14}.
In Ref. \cite{key-14}, Vaz used the Wheeler-DeWitt equation in a
quantum setting to show that the LeMaitre-Tolman-Bondi gravitational
collapse \cite{key-15,key-16,key-17} caused the collapsing matter
to condense naturally on the Schwarzschild surface. This surface,
in turn, happened to be an apparent horizon rather than a real one.
This result allowed Vaz to win the second prize in the Gravity Research
Foundation Essay Competition in 2014. Recently, one of us, CC \cite{key-18},
has shown that these Einstein-Vaz spherical shells can also be obtained
from the gravitational collapse of Oppenheimer and Snyder \cite{key-19}
and turn out to be well-defined quantum objects that obey the Schrödinger
theory in the non-relativistic approximation, and the Klein-Gordon
equation when relativistic corrections are taken into account. We
will return to the Einstein-Vaz shells later in this work since the
fact that their mass spectrum is discrete has important thermodynamical
implications. 

Adapting the analysis in \cite{key-4} we notice that if the dynamical
temperature $T_{E(ECO)}$ of the ECO agrees with the BH dynamical
temperature in Eq. (\ref{eq: Corda Temperature}), then the dynamical
BH entropy (\ref{eq: dynamical BH entropy}) and the probability of
emission in Eq. (\ref{eq: Parikh Correction}) must automatically
agree in the same way. Let us start from the entropy \cite{key-4}.
Considering a body with a large number of degrees of freedom, one
has $dS_{E}=T_{E}dU_{E},$ where the subscripts $E$ indicate the
dynamic values of the various thermodynamic quantities and $dU_{E}$
is the infinitesimal variation of energy. Hence, if the ECO has dynamical
temperature $T_{E(ECO)}$ which matches the BH dynamical temperature
in Eq. (\ref{eq: Corda Temperature}), then one has $d\left(S_{ECO}\right)_{E}=\frac{dU_{E}}{T_{E}}=8\pi M_{E}dM_{E},$
whose integral is 
\begin{equation}
\left(S_{ECO}\right)_{E}=4\pi M_{E}^{2}=4\pi\left(M-\frac{\omega}{2}\right)^{2}=\frac{A_{E}}{4},\label{eq: integrale dinamico}
\end{equation}
where $\omega$ is the ECO emitted energy. The phyisical meaning of
Eq. (\ref{eq: integrale dinamico}) is the following. Assuming that
before the emission of the energy $\omega$ the ECO initial mass is
$M_{i},$ then the ECO initial entropy is 
\begin{equation}
\left(S_{ECO}\right)_{i}=4\pi M_{i}^{2}=\frac{A_{i}}{4},\label{eq: entropia iniziale}
\end{equation}
where the subscript $i$ is assigned to the initial values. After
the emission of the energy $\omega$, the ECO final mass is $M_{f}=M_{i}-\omega.$
Then, the ECO final entropy is 
\begin{equation}
\left(S_{ECO}\right)_{f}=4\pi M_{f}^{2}=4\pi\left(M_{i}-\omega\right)^{2}=\frac{A_{f}}{4}.\label{eq: entropia finale}
\end{equation}
Thus, one understands that the value $\left(S_{ECO}\right)_{E}$ in
Eq. (\ref{eq: integrale dinamico}) represents the value of the entropy
of the ECO during the emission of energy since the mass of the ECO
is decreasing due to the emission of energy $\omega$. Thus, any ECO
with the dynamical temperature $T_{E}$ of the BH, defined by Eq.
(\ref{eq: Corda Temperature}), has the same dynamical entropy $\left(S_{ECO}\right)_{E}$
of the BH.

The second step in adapting the analysis in \cite{key-4} to the BH
dynamical state is to consider the radiation rate, which is connected
with the absorption cross-section as it is required by thermodynamics
\cite{key-1,key-4}. Thus, in thermal approximation, quanta with an
energy range $\left(\omega,\omega+d\omega\right)$ in the spherical
harmonic $Y_{l,m}$ are radiated at a rate 
\begin{equation}
\Gamma_{BH}\left(l,m,\omega\right)d\omega=\frac{P\left(l,m,\omega\right)d\omega}{2\pi\left[\exp(\frac{\omega}{T_{H}})-1\right]}.\label{eq: tasso}
\end{equation}
In this equation, $P\left(l,m,\omega\right)$ is the absorption probability
for an incoming spherical wave having energy $\omega$ in the spherical
harmonic $Y_{l,m}.$ Eq. (\ref{eq: Hilbert dynamical}) permits one
to go beyond the thermal approximation by introducing the \emph{dynamical
rate }
\begin{equation}
\left(\Gamma_{BH}\right)_{E}\left(l,m,\omega\right)d\omega=\frac{P\left(l,m,\omega\right)d\omega}{2\pi\left[\exp(\frac{\omega}{T_{E}})-1\right]}.\label{eq: tasso dinamico}
\end{equation}

For a general emitting source there is no well-defined relationship
between the temperature and the radiation rate. Indeed, the radiation
rate depends on the details of the shape, size and composition of
the source. But, in the particular case of an emitting ECO, the situation
is different \cite{key-4}. If in a region near and outside the horizon
we use the dynamical coordinates
\begin{equation}
s_{E}=\sqrt{8M_{E}\left(r-2M_{E}\right)},\qquad t_{E}=\frac{t}{4M_{E}},\label{eq: coordinate dinamiche di Riindler}
\end{equation}
then the BH dynamical metric of Eq. (\ref{eq: Hilbert dynamical})
becomes the dynamical Rindler space 
\begin{equation}
ds^{2}\approx-s_{E}^{2}dt_{E}+ds_{E}^{2}+dx_{1}^{2}+dx_{2}^{2},\label{eq: dynamical Rindler space}
\end{equation}
where $x_{1}$, $x_{2}$ describe the tangent space to the angular
sphere. Note that the line element on the unit 2-sphere $d\Omega^{2}\equiv\sin^{2}(\theta)d\varphi^{2}+d\theta^{2}$
can be written as $d\Omega^{2}=d\vec{x}\cdotp d\vec{x}=dx_{1}^{2}+dx_{2}^{2}$,
where $d\vec{x}\equiv dx_{1}\hat{x}_{1}+dx_{2}\hat{x}_{2}$ with $dx_{1}\hat{x}_{1}\equiv d\theta\hat{e}_{\theta}$
and $dx_{2}\hat{x}_{2}\equiv\sin(\theta)d\varphi\hat{e}_{\varphi}$.
The orthonormal vectors $\hat{e}_{\theta}\equiv\frac{\partial_{\theta}\vec{r}}{\left\Vert \partial_{\theta}\vec{r}\right\Vert }=\left(\cos(\theta)\cos(\varphi),\cos(\theta)\sin(\varphi),-\sin(\theta)\right)$
and $\hat{e}_{\varphi}\equiv\frac{\partial_{\varphi}\vec{r}}{\left\Vert \partial_{\varphi}\vec{r}\right\Vert }=\left(-\sin(\varphi),\cos(\varphi),0\right)$
are the tangent vectors that span the tangent space to the unit 2-sphere
at the point $P=\left(r,\:\theta,\:\varphi\right)$ with vector position
$\vec{r}\equiv\left(r\sin(\theta)\cos(\varphi),r\sin(\theta)\sin(\varphi),r\cos(\theta)\right)$. 

In relativistic quantum mechanics, the particle content of a single
quantum state in two distinct reference frames is generally different
\cite{key-32}. This difference is due to the fact that distinct reference
frames are specified by distinct notions of time. Therefore, the concept
of frequency, with particles viewed as excitations of positive frequency
modes, is observer-dependent {[}20--22{]}. A remarkable effect that
emerges from this observer-dependence is the fact that a vacuum in
an inertial frame becomes a thermal state to an accelerating observer
(this is the so-called Unruh effect \cite{key-22}). For a clear presentation
on how to transform a quantum state that is known in terms of Minkowskian
(Rindler) excitations into Rindler (Minkowskian) space, we refer to
Ref. \cite{key-23}. The Boulware vacuum denotes the vacuum with respect
to the Schwarzschild time. In the Schwarzschild space, the Boulware
vacuum is stable and, in addition, is the lowest-energy state \cite{key-24,key-25}.
In BH semiclassical physics, the region near the horizon is locally
Minkowskian, with its quantum state being close to the Minkowskian
vacuum in \cite{key-4}. The line element of Eq. (\ref{eq: dynamical Rindler space})
concerns the right Rindler wedge of this Minkowskian zone. In a Rindler
frame, the Minkowskian vacuum results to be the Rindler vacuum plus
a series of non-strictly thermal excitations. In a BH framework, the
Rindler vacuum is the Boulware vacuum with a temperature $T(s)\approx\frac{2\pi}{s},$
where $s$ is the proper distance outside the horizon. One notes that,
unlike the strictly thermal scenario in the case where one considers
deviations from thermality, the horizon cannot be fixed but is subject
to quantum oscillations due to the BH back reaction and variable geometry.
This means that the temperature $T(s)$ must be replaced by the dynamical
temperature $T_{E}(s_{E})\approx\frac{2\pi}{s_{E}}$ ($s_{E}$ being
the dynamical proper distance outside the horizon) which will be associated
with the BH dynamical temperature of Eq. (\ref{eq: Corda Temperature}),
blueshifted to the local orthonormal frame at $s_{E}$, instead of
the traditional Hawking temperature in Eq. (\ref{eq: Hawking temperature}): 

\begin{equation}
\frac{T_{E}(\omega)}{\sqrt{-g_{tt}}}=\frac{1}{2\pi s_{E}}.\label{eq: blueshifted}
\end{equation}
The left-hand-side of Eq. (\ref{eq: blueshifted}) is the usual gravitational
shift law in Eq. (25.26) of \cite{key-27}, adapted to the gravitational
blueshift of the dynamical temperature $T_{E}(\omega)$. Moreover,
$g_{tt}$ is given in Eq. (\ref{eq: Hilbert dynamical}) and equals
$-\left(1-\frac{2M_{E}}{r}\right).$ Here are some details of the
computation. The gravitational blueshift of the dynamical temperature
$T_{E}$ in spherical symmetry is given by $T_{E}\rightarrow\frac{T_{E}}{\sqrt{-g_{tt}}}$
\cite{key-27}. From the first relation in Eq. (\ref{eq: coordinate dinamiche di Riindler})
one gets 
\begin{equation}
r=\frac{s_{E}^{2}}{8M_{E}}+2M_{E},\label{eq: details}
\end{equation}
where $2M_{E}=2M-\omega$ from Eq. (\ref{eq: effective quantities emission}).
Then, we have
\begin{equation}
\frac{T_{E}(\omega)}{\sqrt{-g_{tt}}}=\frac{1}{4\pi(2M-\omega)}\frac{1}{\sqrt{1-\frac{1}{1+\frac{s_{E}^{2}}{16M_{E}^{2}}}}}\simeq\frac{1}{8\pi M_{E}}\frac{4M_{E}}{s_{E}}=\frac{1}{2\pi s_{E}}.\label{eq: dettagli}
\end{equation}
In the derivation of the Hawking radiation in Schwarzschild coordinates
for a scalar field $\hat{\Phi}$ satisfying $\square\hat{\Phi}=0,$
one expands the scalar field in modes in the external dynamical region
$r>2M_{E}$ as \cite{key-4} 
\begin{equation}
\begin{array}{c}
\hat{\Phi}=\sum_{l,m,k}[\hat{a}_{l,m,k}f_{l,m,k}(r)Y_{l,m}\left(\theta,\phi\right)\exp\left(-i\omega_{l,m,k}t\right)\\
\\
+\hat{a}_{l,m,k}^{\dagger}f_{l,m,k}^{*}(r)Y_{l,m}^{*}\left(\theta,\phi\right)\exp\left(i\omega_{l,m,k}t\right)].
\end{array}\label{eq: fi}
\end{equation}
The quantity $f_{l,m,k}(r)$ is the effective potential to which the
modes of Eq. (\ref{eq: fi}) are subject. The quantities $\hat{a}_{l,m,k}$
and $\hat{a}_{l,m,k}^{\dagger}$ are the creation and annihilation
operators, respectively. One has a barrier that separates the region
near the horizon from the region at infinity. Since the horizon is
not fixed but is oscillating due to successive Hawking quanta emissions,
Eq. (\ref{eq: Hilbert dynamical}) allows us to use the dynamical
horizon in Eq. (\ref{eq: effective quantities emission}). The dynamical
rate in Eq. (\ref{eq: tasso dinamico}) is obtained as follows \cite{key-4}:
\begin{enumerate}
\item In the region near the dynamical horizon, the local Minkowski vacuum
happens to be the local Boulware vacuum plus a quasi-thermal gas of
excitations $\hat{a}_{l,m,k}^{\dagger}$ over the Boulware vacuum.
The quasi-thermal gas turns out to have a dynamical temperature given
by Eq. (\ref{eq: blueshifted}) at a dynamical proper distance $s_{E}$
from the dynamical horizon. 
\item These modes overcome the barrier due to the tunneling effect, reaching
infinity with a probability $P\left(l,m,k\right)$ and, therefore,
generating the (not strictly thermal) Hawking radiation.
\item Since the probability of tunnelling is symmetric for incoming and
outgoing modes, this probability $P\left(l,m,k\right)$ equals the
probability that the same mode is absorbed by the BH. This gives rise
to the dynamical rate of emission in Eq. (\ref{eq: tasso dinamico}).
\end{enumerate}
Following \cite{key-4}, one supposes to replace the horizon by an
oscillating ECO (for example, an Einstein-Vaz spherical shell) having
the same dynamical BH temperature. Consequently, the ECO's surface
should populate the region just outside this surface with particles
of the field $\hat{\Phi}$ with the same dynamical temperature as
in Eq. (\ref{eq: blueshifted}). Since the dynamical metric in the
region outside the ECO is the same as the BH dynamical metric associated
to the same dynamical mass $M_{E}$ (in both cases it is the dynamical
Schwarzschild metric in Eq. (\ref{eq: Hilbert dynamical esplicita})),
the potential barrier is the same as the barrier for the BH. Hence,
the ECO dynamical radiation rate is equal to the BH dynamical radiation
rate in Eq. (\ref{eq: tasso dinamico}). The key point that allows
one to achieve this result is that the wavelengths of the quasi-thermal
particles near the dynamical Schwarzschild surface are very short
compared to the length scale of the extension of the effective gravitational
potential \cite{key-4}. This is due to the third of the properties
that define the ECO under Eq. (\ref{eq: dynamical BH entropy}) and
allows one to separate the quasi-thermal physics near the Schwarzschild
surface from the dynamics of tunneling through the barrier \cite{key-4}.

From the discussion made so far, it emerges that if an ECO had the
same dynamical temperature as a BH, then it must also have the same
dynamical entropy along with the same dynamical emission rate. So,
to obtain what is proposed in this paper, i.e. to show that the BH
thermodynamics is equal to the ECO thermodynamics even in a scenario
that is not exactly thermal, one must demonstrate that the BH and
the ECO have the same dynamical temperature. Following \cite{key-4}
again, let us consider the \emph{energy densities} near the dynamical
gravitational radius $r\approx r_{E}=2M_{E}.$ Since for a BH the
region around the horizon is a Minkowski vacuum, then, to leading
order, the expectation value of the total energy density must be zero
because the geometry is smooth at the horizon, and one finds no flux
into or out of the horizon. Note that $\rho_{T}\equiv-<T_{0}^{0}>$
, with $T_{0}^{0}$ being the $00-$component of the stress-energy
tensor $T^{\mu\nu}.$ For more details on the calculation of (renormalized)
vacuum expectation values of $T^{\mu\nu},$ we refer to \cite{key-28}.
In the coordinates of Eq. (\ref{eq: Hilbert dynamical}), the energy
density of excitations at dynamical temperature (\ref{eq: dettagli})
for the Unruh vacuum is \cite{key-29} $<0_{U}|T_{0}^{0}|0_{U}>=\rho_{U}=\frac{1}{48\pi^{2}s_{E}^{4}}.$
This energy density is added to the negative Casimir energy density
of the Rindler vacuum, which is exactly its opposite, $<0_{R}|T_{0}^{0}|0_{R}>=\rho_{R}=-\frac{1}{48\pi^{2}s_{E}^{4}},$
so that the total energy density is zero. On the other hand, an ECO
at zero dynamical temperature has approximately the same negative
Casimir energy \cite{key-4} $\rho_{ECO,R}\simeq-\frac{1}{48\pi^{2}s_{E}^{4}},$
at $r\gtrsim R_{E(ECO)},$ where $R_{E(ECO)}$ is the dynamical oscillating
radius of the ECO (in the case of the Einstein-Vaz shell this is exactly
equal to the dynamic BH gravitational radius of Eq. (\ref{eq: effective quantities emission})
and obeys a Schrödinger equation completely analogous to that of the
s-states of the hydrogen atom \cite{key-18}). This can be understood
by expanding the scalar field $\hat{\Phi}$ in modes in the background
of the ECO as \cite{key-4} 
\begin{equation}
\begin{array}{c}
\hat{\Phi}=\sum_{l,m,k}[\hat{b}_{l,m,k}g_{l,m,k}(r)Y_{l,m}\left(\theta,\phi\right)\exp\left(-i\omega_{l,m,k}t\right)\\
\\
+\hat{b}_{l,m,k}^{\dagger}g_{l,m,k}^{*}(r)Y_{l,m}^{*}\left(\theta,\phi\right)\exp\left(i\omega_{l,m,k}t\right)],
\end{array}\label{eq: scalar field expanded}
\end{equation}
and defining the vacuum state for the ECO $|0>_{ECO}$ via the condition
$\hat{b}_{l,m,k}|0>_{ECO}=0$ for all $l,m,k.$ As the wavemode in
the ECO has a large number of oscillations in the region $r>R_{E(ECO)}$
(of order of $\log\left(\frac{GM}{l_{p}}\right)$ in standard units,
with $G$ and $l_{p}$ being the gravitational constant and the Planck
lenght, respectively \cite{key-4}), one makes well-defined wavepackets
out of these modes to study the physics at any point $r\gtrsim R_{E(ECO)}$
without needing to know the form of modes in $r<R_{E(ECO)}.$ By using
the modes in Eq. (\ref{eq: fi}), Candelas \cite{key-28} derived
the BH Casimir energy density in the Boulware vacuum as 
\begin{equation}
\rho_{B}=-\frac{1}{48\pi^{2}s^{4}},\label{eq: Casimir BH}
\end{equation}
where $s$ is the standard proper distance outside the horizon. Mathur
and Mehta \cite{key-4} obtained the same result, to an excellent
approximation, for the ECO Casimir energy density in the Boulware
vacuum as 
\begin{equation}
\rho_{ECO,B}\simeq-\frac{1}{48\pi^{2}s^{4}}.\label{eq: Casimir ECO}
\end{equation}
Remembering that the external geometry of the ECO is the same as that
of the BH (in both cases it is the Schwarzshild one), using the dynamical
metric in Eq. (\ref{eq: Hilbert dynamical}), one gets finally 
\begin{equation}
\rho_{ECO,B}\simeq\rho_{B}=-\frac{1}{48\pi^{2}s_{E}^{4}},\label{eq: Casimir ECO BH}
\end{equation}
where the dynamical proper distance outside the horizon $s_{E}$ replaced
$s$. 

Now, again adapting the analysis in {[}4{]}, one assumes that the
dynamic temperature of the ECO is zero so that the ECO has a corresponding
quantum ket $|0>_{ECO}$. The meaning of the Casimir density energy
in Eq. (\ref{eq: Casimir ECO BH}) is the presence of negative energy
in the semiclassical region just outside of the oscillating $R_{E(ECO)}.$
One considers the energy of a thin spherical shell having an oscillating,
dynamical internal radius, in terms of its dynamical proper distance,
$\bar{s}_{E}=\mu l_{P}=\mu$ , since $l_{P}\equiv\sqrt{\frac{\hbar G}{c^{3}}}=1$
in Planck units, outside $r_{E}=2M_{E}$. The energy of the shell
is 
\begin{equation}
E_{shell}\approx4\pi\left(2M_{E}\right)^{2}\int_{\bar{s}_{E}}^{\infty}\rho_{ECO,B}(s_{E})\sqrt{-g_{tt}(s_{E})}ds_{E},\label{eq: E shell zero}
\end{equation}
with 
\begin{equation}
\sqrt{-g_{tt}(s_{E})}=2\pi s_{E}T_{E}(\omega)=\frac{s_{E}}{4M_{E}}\label{eq:  rottura}
\end{equation}
from Eq (\ref{eq: blueshifted}). Then, one gets 
\begin{equation}
E_{shell}\approx16\pi M_{E}^{2}\int_{\bar{s}_{E}}^{\infty}\left(-\frac{1}{48\pi^{2}s_{E}^{4}}\right)\left(\frac{s_{E}}{4M_{E}}\right)ds_{E}=-\frac{M_{E}}{24\pi\bar{s}_{E}^{2}}=-\left(\frac{1}{24\pi\mu^{2}}\right)M_{E}=-XM_{E},\label{eq: E shell}
\end{equation}
with $X\equiv\frac{1}{24\pi\mu^{2}}.$ One now takes $\mu=\mathcal{O}(1)$
and assumes that $R_{E(ECO)}$ is the Planck length $r_{E}=2M_{E}$
and we can imagine that $\bar{s}_{E}$ is five times the Planck length
\cite{key-4}. So, one has that also $X=\mathcal{O}(1).$ In this
case, we note that the ECO under consideration has a problem \cite{key-4}.
The dynamical mass as seen from infinity is $M_{E}.$ Hence, in the
region $r<R_{E(ECO)}$ the effective mass should be 
\begin{equation}
M_{E}(r<R_{E(ECO)})=M_{E}+XM_{E}=\left(1+X\right)M_{E}.\label{eq: massa dinamica interna ECO}
\end{equation}
But this dynamical mass implies a dynamical horizon outside $R_{E(ECO)},$
which, in turn, implies a violation of the 4th ECO property associated
to Eq. (\ref{eq: Casualit=0000E0}). Consequently, an ECO having zero
dynamical temperature cannot exist. A similar situation occurs for
every ECO dynamical temperature $T_{E(ECO)}$ lower than the BH dynamical
temperature $T_{E}$. The state corresponding to the dynamical temperature
of the ECO is obtained by adding a quasi-thermal bath of excitations
$\hat{b}_{l,m,k}$ to the vaccum state $|0>_{ECO}.$ Then, the local
dynamical temperature near the dynamical horizon reads 
\begin{equation}
T_{E(ECO)}(s_{E})=\frac{2\pi}{s_{E}}\frac{T_{E(ECO)}}{T_{E}}.\label{eq: relazione temperature}
\end{equation}
Thus, one finds the energy in the shell region as 
\begin{equation}
E_{shell}\approx-\frac{M_{E}}{24\pi\bar{s}_{E}^{2}}\left[1-\left(\frac{T_{E(ECO)}}{T_{E}}\right)^{4}\right]=-\frac{1}{24\pi\mu^{2}}\left[1-\left(\frac{T_{E(ECO)}}{T_{E}}\right)^{4}\right]M_{E}=-X(T_{E})M_{E},\label{eq: E shell 2}
\end{equation}
where one still has $X(T_{E})=\mathcal{O}(1)$ with $X(T_{E})\equiv\frac{1}{24\pi\mu^{2}}\left[1-\left(\frac{T_{E(ECO)}}{T_{E}}\right)^{4}\right]$.
Therefore, even in this case the ECO cannot exist. One now wonders
what happens if it is $T_{E(ECO)}>T_{E}$. Still following the approach
in \cite{key-4} one has no viable solution to the Tolman-Oppenheimer-Volkoff
equation which describes the quasi- thermal radiation in the region
outside $R_{E(ECO)}.$ In fact, the third property that specifies
the ECO implies that the redshift reaches $\mathcal{O}(M_{E})$ at
$R_{E(ECO)}.$ One specifies a parameter $u_{E}$ which allows one
to characterize the redshift as $\frac{4M_{E}}{u_{E}}.$ So, it must
be $u_{E}$ of order of the Planck length. Following the second property
that defines the ECO, in the region outside $R_{E(ECO)}$ standard
semi-classical physics works. Hence, in this region the local dynamical
temperature is $\frac{4M_{E}}{u_{E}}T_{E}.$ Then, the local energy
density in this region outside $R_{E(ECO)}$ arises from this dynamical
temperature and the energy in a shell outside $R_{E(ECO)}$ is 
\begin{equation}
E_{shell}\approx\frac{1}{24\pi u_{E}^{2}}\left[\left(\frac{T_{E(ECO)}}{T_{E}}\right)^{4}-1\right]M_{E}=Y(T_{E})M_{E},\label{eq: E shell 2-1}
\end{equation}
with $Y(T_{E})\sim1.$ Thus, one finds a violation of the third property
that defines the ECO, which required that the dynamical mass outside
$R_{E(ECO)}$ be $\mathcal{O}(M_{E})$, that is, negligible compared
to $M_{E}$. Hence, exactly as in the previous cases, this ECO cannot
exist in this case as well.

For a more detailed discussion on the fact that any ECO has, to leading
order, the same thermodynamic properties as the semiclassical BH of
the same mass, we refer to the recent work by Mathur and Mehta in
Ref. \cite{key-33}.

\section*{Concluding remarks}

The concepts of mass, entropy, and information lead naturally to thermodynamic
arguments \cite{key-31}. A prominent example of such an occurrence
is represented by the important work of Mathur and Mehta on the universality
of BH thermodynamics \cite{key-4}. In Ref. \cite{key-4} Mathur and
Mehta showed that any ECO must have the same BH thermodynamic properties
regardless of whether the ECO possesses an event horizon or not. In
this paper, we use the concept of BH dynamical state developed in
Ref. \cite{key-6} to extend the analysis of Mathur and Mehta in Ref.
\cite{key-4} to the case where the BH radiation spectrum is not exactly
thermal. Our results are consistent with the Einstein-Vaz shell approach
developed in Refs. \cite{key-14,key-18,key-30}. According this approach,
in both the LeMaitre-Tolman-Bondi and the Oppenheimer-Snyder gravitational
collapses, the collapsing matter condenses naturally on the Schwarzschild
surface which, in turn, results to be an apparent horizon rather than
a real one. Interestingly, the Einstein-Vaz shells are horizon-sized
quantum objects that obey the Schrödinger and Klein-Gordon theories
\cite{key-18} and radiate from their surface like normal bodies having
a discrete mass spectrum. Therefore, there is no information loss
paradox. Indeed, the time evolution of such Einstein-Vaz shells is
governed by a time-dependent Schrödinger equation which guarantees
the unitarity of the evaporation process assumed to be not strictly
thermal \cite{key-30}, in consistent agreement with the shells discrete
mass spectrum.

For completeness, it is right to emphasize the following \cite{key-34}.
The corrections proposed in this letter, compared to the pure thermal
approximation in \cite{key-4}, are likely to be very small, since
the energy of each emission in the BH is so small. In the arguments
relating ECOs to BHs, the agreement of thermodynamic quantities was
only ``to leading order'{}'' which means that the agreement gets
better and better as the ECO radius gets closer and closer to the
BH horizon radius. Thus it would be good to know if the small changes
expected from the modification studied in this paper can be noticed
above the approximation inherent in the comparison of the ECO and
the BH. This is certainly an interesting idea to carry forward in
subsequent works.

\section*{Acknowledgments}

The Authors thank an anonymous Referee for helpful comments.

\end{document}